\documentstyle[prb,preprint,aps]{revtex}

\tighten

\begin{document}
\draft

\title{From Cooper Pairs to Composite Bosons:\\
A Generalized RPA Analysis of Collective Excitations}


\author{
          Lotfi Belkhir\\[10pt]
          Department of Physics,\\
          State University of New York, \\
          Stony Brook, New York 11794-3800 \\[15pt]
          Mohit Randeria\\[10pt]
          Argonne National Laboratory, MSD 223 \\
          9700 S.~Cass Ave., Argonne, IL 60439 \\
         }

\maketitle

\begin{abstract}
We study the evolution of the ground state and the
excitation spectrum of the two and three dimensional attractive
(negative-$U$) Hubbard model as the system evolves
from a Cooper pair regime for $U \ll t$,
to a composite boson regime for $U \gg t$.
Our work is motivated by the observation that the
high temperature superconductors, with their
short coherence lengths and unusual normal state properties,
may be in an intermediate coupling regime between these two limits.
A mean field analysis of pairing, suitably generalized to account for
a shift in the chemical potential, is known to be able to describe the
ground state crossover as a function of $U/t$.
We compute the collective mode spectrum using a generalized
random phase approximation (RPA) analysis within the
equations of motion formalism. We find a smooth evolution of the
Anderson mode for weak coupling into the Bogoliubov sound mode for
hard core bosons. We then include a long-range Coulomb interaction
and show that it leads to a plasmon which again evolves smoothly from
weak to strong coupling.

\end{abstract}

\pacs{PACS numbers: 74.20.Fg, 71.45.Gm}

\mediumtext

\section{Introduction}

Since their discovery by Bednorz and Muller\cite{hitc}, six
years ago, the cuprate superconductors have consistently shown
appreciable deviations\cite{losalamos}, both in their superconducting and
normal state properties, from the conventional superconducting metals.
One of the most remarkable characteristics of the high-Tc
materials is their short coherence length which is a few times the
lattice spacing so that the pairs of fermions are weakly
overlapping in real space. This is in marked
contrast with the BCS superconductors where
the pair size is orders of magnitude larger then the lattice spacing,
and the number of fermions within a Cooper pair is very large.

The problem of the crossover from a BCS state with Cooper pairs
to a condensate of composite bosons was first
considered at $T=0$ by Leggett\cite{ajl}
and further analyzed at finite temperatures
by Nozieres and Schmitt-Rink\cite{nsr}.
These ideas were taken up again
in the context of the high $T_c$ superconductors
where it was suggested\cite{rds} that these systems might
properly be described as being in an intermediate regime
between the BCS limit and the Bose limit.
Very recently, the normal state in the intermediate
coupling regime has been investigated\cite{rtms} by Monte Carlo
simulations of the 2D attractive Hubbard model,
with a view to studying deviations from Fermi liquid behavior.
It was found\cite{rtms} that a degenerate Fermi system shows
anomalous spin-correlations above $T_c$, providing a natural
qualitative explanation of the spin-gap behavior observed\cite{mt} in
the Knight shift and NMR relaxation rate
of several high-$T_c$ systems.

In this paper we study the collective excitations in the attractive Hubbard
model at $T=0$ as the ground state evolves from the BCS to the Bose regime.
A study of the collective modes as a function of coupling
is important for at least two reasons.
The critical temperature in the BCS
weak coupling regime, as well as the
thermodynamic properties, are governed by the pair breaking
excitations with an  energy gap $\Delta$ which is exponentially small.
However, in the strong coupling regime of composite bosons
the energy gap becomes very large and pair breaking is no
longer possible. $T_c$ is then controlled by the center of mass
motion of the pairs, or in other words by the {\sl
collective modes}. In the intermediate regime, the thermodynamic
properties will be controlled by a combination of pair breaking and
collective modes. A study of the $T=0$ collective modes is then a first step
towards understanding this complicated regime.

A second reason for studying the collective modes is that, at least in
the Bose regime, we know from Bogoliubov theory that their spectrum
at long wavelengths depends in an essential way on the repulsive
interaction between the bosons. While the crossover analyses, refered
to above, were able to access the Bose regime while working with
the constituent fermions,
it is not apriori clear how the effective
interactions between the pairs are accounted for in these treatments.

We study the collective mode spectrum at $T=0$ using a generalized
RPA formulation. We adapt the analysis of
Anderson\cite{pwa} and others\cite{rick,ak,bs,parks}, initially
applied to weak coupling superconductivity, and show that with some
modifications it is capable of describing the evolution of the collective
mode for {\it all} values of $U/t$. While one might not expect the RPA to
be valid in the strong coupling limit, we show that we recover the well
known Bogoliubov result\cite{bog}
for the sound velocity of a repulsive Bose gas
in the $U \gg t$ limit. Using the RPA as an interpolation
scheme we find a smooth evolution of the collective
mode spectrum in both two and three dimensions.

We extend the anaysis to
charged superconductors in two and three dimensions.
In weak coupling we find that the sound mode for neutral system is
pushed up\cite{pwa} to the plasma frequency in 3D;
in 2D the plasmon has a $\sqrt{q}$ dispersion.
We find that the plasmon evolves smoothly
as a function of the attraction, and in the strong coupling, dense system limit
we recover exactly the known plasma frequency for a dense charged Bose
gas\cite{foldy}.

Some of the results of this paper have been
presented without detailed derivation in
a previous Rapid Communication\cite{br}.
Related results have been obtained by
several authors\cite{others,griffin} using different techniques.
The detailed presentation given here is nevertheless of
some interest, especially since we derive analytical results
for the BCS and Bose limiting cases.
Recently, the RPA has also been successfully applied to other
crossover problems, for example, the evolution from
itinerant to local moment antiferromagnetism\cite{swz},
and excitonic collective modes in a Bose condensed
electron-hole gas\cite{excitons}.

\section{Mean Field Analysis}

\noindent
Our starting point is the single band attractive Hubbard model
on a $d$-dimensional hypercubical lattice: we will focus on
the 2D and 3D cases. The Hamiltonian
\begin{equation}
H = -t\sum_{ij\sigma}c^{+}_{i\sigma}c_{j\sigma} - U\sum_{i}n_{i\uparrow}
n_{i\downarrow} - \mu\sum_{i}\left(n_{i\uparrow} + n_{i\downarrow}\right)
\end{equation}
\noindent
may be rewritten in momentum space as
\begin{equation}\label{hamil}
H = \sum_{k,\sigma} (\varepsilon_{k}-\mu)c^{+}_{k\sigma}c_{k\sigma} -
\frac{U}{M}\sum_{kk^{\prime}q}c^{+}_{k+q\uparrow}
c^{+}_{k^{\prime}-q\downarrow}c_{k^{\prime}\downarrow}c_{k\uparrow}
\end{equation}
where $\varepsilon_{k}=-2t\sum_{i=1}^{d}cos(k_{i}a)$.
The chemical potential $\mu $ will be adjusted to obtain
the required band filling $f = N/2M$, where $N$ is the average
number of electrons and $M$ the total number of sites.
We will study the model at $T=0$ for arbitrary $U/t > 0$ and
$0\le f < 1/2$ where the ground state is expected to have
superconducting off-diagonal long range order.
We will not discuss the competition
between superconductivity and CDW ordering\cite{montecarlo}
at half-filling $f = 1/2$.

We quickly review the mean field analysis, mainly to establish notation.
The BCS ``reduced'' Hamiltonian $H_{\rm bcs}$
is that part of (\ref{hamil}) which describes
the interaction between pairs with zero center-of-mass momentum.
$H_{\rm bcs}$ is thus obtained by retaining
only the $k^{\sl{Phys. Rev.},ime}=-k$ piece of the second term of
(\ref{hamil}).
The BCS-Bogoliubov solution consists of
determining the eigenoperators
$\gamma^{+}_{k\sigma}$ and $\gamma_{k\sigma}$ of $H_{\rm bcs}$.
These operators define both the BCS ground state,
via $\gamma_{k\sigma}|\Phi_{0}\rangle = 0$,
and the single-particle excitations
$\gamma^{+}_{k\sigma}|\Phi_{0}\rangle$ with energy $E_{k}$.

The first step is a linearization
of the equations of motion for $c_{k\sigma}$
and $c^{+}_{k\sigma}$ with respect to
$\vert \Phi_0 \rangle$ where
$\langle c^{+}_{k\sigma}c_{k\sigma}\rangle$
and $\langle c^{+}_{k\uparrow}c^{+}_{-k\downarrow}\rangle$ are
non-vanishing. We must retain the
Hartree shift in the chemical potential
\begin{equation}\label{hartree}
\tilde{\mu} = \mu + fU
\end{equation}
since it plays a crucial role for large $U$ (see below).
The second step is to diagonalize the linearized equations
via the Bogoliubov transformation:
$c_{k \uparrow} = u_{k}\gamma_{k 0} + v_{k}\gamma^{+}_{k 1}$ and
$c^{+}_{-k \downarrow} = - v_{k} \gamma_{k 0} + u_{k}\gamma^{+}_{k 1}$.
Here $u_k^2 = 1 - v_k^2 = {1\over2}\left(1 + \xi_k/E_k \right)$,
with $\xi_k = \varepsilon_k - \tilde{\mu}$, and
the order parameter and the quasiparticle excitation energy are given by
$\Delta = U\sum_k u_k v_k$, and
$E_{k} = (\xi_{k}^{2} + \Delta^{2})^{1/2}$ respectively.

Self-consistency is achieved by demanding that,
for each value of the coupling $U/t$ and filling $f$,
$\Delta$ and the chemical potential $\tilde{\mu}$
satisfy the gap equation
\begin{equation}\label{gapeqn}
1  = \frac{U}{2M}\sum_{k}\frac{1}
{[\xi_{k}^{2}+\Delta^{2}]^{1/2}}
\end{equation}
which is familiar from BCS theory, and the number equation
\begin{equation}\label{numbereqn}
f  = \frac{1}{2M}\sum_{k}\left(1 - \frac{\xi_{k}}
{[\xi_{k}^{2}+\Delta^{2}]^{1/2}}\right) .
\end{equation}
The latter, which simply follows from
$2\sum_{k}v_k^2 = N$, is trivially solved in the
BCS limit to obtain $\mu \simeq \varepsilon_F$,
the noninteracting result.
As the coupling grows, however, the occupation probabality in momentum space
$n_k = 2v_k^2$ broadens significantly, corresponding to the formation
of tightly bound pairs. As a result the chemical potential is strongly
affected\cite{ajl} by the interactions
(in addition to the trivial Hartree shift (\ref{hartree}).

The gap and number equations may be solved analytically in the weak and
strong coupling limits\cite{ajl,nsr}.
(In the continuum limit in 2D
an exact analytical solution is possible for
{\it arbitrary} couplings\cite{rds}).
For $U/t \ll 1$ the chemical potential is at $\varepsilon_F$,
the pair size is much larger than the lattice spacing, as expected
of a BCS ground state, and the energy gap has the usual essential
singularity in $U/t$.
In the opposite limit of $U/t \gg 1$, the pairs are on-site,
the chemical
potential is one-half the pair binding energy, and the ground state is
a condensate of composite bosons.
The expressions for $\tilde{\mu}$ and $\Delta$ in the strong coupling
regime were initially given by Nozieres and Schmitt-Rink\cite{nsr};
\begin{equation}
\tilde{\mu} = \frac{U}{2}(2f-1);~~~~~~
\Delta = U\sqrt{f(1-f)}
\end{equation}
There is a smooth crossover between these rather different limits, as can be
seen by a numerical solution of (\ref{gapeqn}) and (\ref{numbereqn}).
The solutions in 3D case are given in ref.~\cite{nsr} and the
2D results are shown in Fig.~1a and b where we plot
$\Delta$ and $\mu$ as a function of $U$ for various fillings $f$.
The gap to single-particle excitations is defined by
\begin{equation}
E_{\rm gap} =
{\rm min}\sqrt{\left(\varepsilon_k - \tilde\mu\right)^2 + \Delta^2}
\end{equation}
where the minimum is to be found within the band, i.e.,
$\varepsilon_k > -2dt$. We thus find $E_{\rm gap} = \Delta$
provided $\tilde\mu$ lies within the band. However,
once $\tilde{\mu}$ is below the bottom of the band\cite{ajl,rds},
we find
$E_{\rm gap} = \left(
\left(- 2dt + |\tilde\mu|\right)^2 + \Delta^2\right)^{1/2}$.

To understand how a BCS-like analysis is able to describe
a condensate of composite bosons for $U \gg t$, note that
the (un-normalized) BCS ground state can be written as
$\vert \Phi_0 \rangle = \left(\sum_k \varphi_k
c^{+}_{k\uparrow}c^{+}_{-k\downarrow} \right)^{N/2} \vert {\rm vac} \rangle$,
with $\varphi_k = u_k/v_k$.
Variationally, the mean field solution corresponds to an
optimal choice of the internal pair wave-function $\varphi$,
which in the strong coupling limit is an on-site singlet.
The Pauli exclusion principle for the constituent fermions
gives rise to a hard core repulsion for these composite bosons.
As shown below, this will have an important effect on the
collective mode spectrum.

\vspace{0.5cm}
\section{ Generalized RPA}

At the mean field level we treated only pairs with zero center of
mass momentum, by retaining only the $H_{\rm bcs}$
part of the full Hamiltonian $H = H_{\rm bcs} + H_{\rm int}$.
In this section we shall treat within a generaized RPA
the fluctuations introduced by $H_{int}$
which describes the interaction between the Bogoliubov
quasiparticles.

The RPA can be implemented in a variety of ways to study the
collective excitations above the superconducting ground state.
We use here the linearized
equations of motion  method which, in our view, is physically
transparent and has a certain intuitive appeal.
This method had been originally developed
for the normal fermi liquid state by Bohm and Pines\cite{bp}, and adapted
to the superconducting state by Anderson\cite{pwa}.

To determine the collective mode spectrum we study the time evolution of
density fluctuations $\rho_{k\sigma}(q)=c_{k+q\sigma}^{+}c_{k\sigma}$.
Its equation of motion
is coupled to that of pairs with a finite center of mass momentum
$b_{k\uparrow}^{+}(q)=c_{k+q\uparrow}^{+}c_{-k\downarrow}^{+}$ and
$b_{k\uparrow}(q)=c_{-k-q\downarrow}c_{k\uparrow}$,
resulting from the particle-hole mixing due to the condensate.
The first step in the generalized RPA is to linearize the equations of
motion of these operators and the second to diagonalize them
by finding the appropriate eigenoperators for the collective coordinates.

Before proceeding with the calculation,
it may be useful to write the terms appearing in the linearized
equations of motion as diagrams. In Fig.~2 we show
the first order processes involved in the full linearized
equations of motion for $c_{k+q\uparrow}^{+}c_{-k\downarrow}^{+}$
(pair or p-p channel) and
in Fig.~3 those for $c_{k+q\sigma}^{+}c_{k\sigma}$ (density fluctuation,
or p-h channel).
Note that no exchange scattering is allowed by the on-site (attractive)
Hubbard interaction.
In Fig.~2, (2a) is the
usual particle-particle scattering vertex, and (2b) is just the Hartree
self-energy term which renormalizes the chemical potential.
The existence of the condensate (a non-zero expectation value of
$<b_{k\sigma}^{+}(q=0)>$) mixes the p-p and p-h channels as shown
in (2c) and (2d).

We now plunge into the rather lengthy
algebra of the equations of motion method.
We use a derivation based on the very clear presentation of Bardasis
and Schrieffer\cite{bs}, adapting it to the lattice model,
and retaining terms which allow us to work at arbitrary $U$.
The weak coupling BCS analysis is considerably simplified by
an approximate particle-hole symmetry arising due to the fact that
only fermions in a thin shell which
is symmetric about the Fermi surface are affected by the pairing.
For arbitrary $U$ one no longer
has such a p-h symmetry which makes the calculation rather more
complicated.

We look at the time evolution of
$\gamma_{k+q,\sigma}^{+}\gamma_{k,\sigma}$,
$\gamma_{k+q,0}^{+}\gamma_{k,1}^{+}$, and $\gamma_{k+q,1}\gamma_{k,0}$,
(instead of working with bilinear products of $c$ and $c^{+}$).
Using the notation
\begin{equation}
\label{qp:energy}
E_{p,q}=E_{p+q}+E_{p}
\end{equation}
the Anderson-Rickayzen equations are given by
the following commutation relations:
\begin{eqnarray}
\left[H , \gamma_{p+q,0}^{+}\gamma_{p,1}^{+}\right] &=& E_{p,q}
\gamma_{p+q,0}^{+} \gamma_{p,1}^{+} - \frac{U}{2M}\,m(p,q)
\sum_{k} \,m(k,q) \left(\gamma_{k+q,0}^{+}\gamma_{k,1}^{+} +
       \gamma_{k+q,1}\gamma_{k,0}\right) \nonumber \\
  & & -\frac{U}{2M}\,l(p,q) \sum_{k}\,l(k,q)\left(
\gamma_{k+q,0}^{+} \gamma_{k,1}^{+} - \gamma_{k+q,1}
\gamma_{k,0}\right) \nonumber \\
   & &  - \frac{U}{2M}\,n(p,q)\sum_{k}\,n(k,q)
\left(\gamma_{k+q,0}^{+}\gamma_{k,1}^{+} + \gamma_{k+q,1}
\gamma_{k,0} \right)       \label{eq: eqmotion1} \\
\left[H , \gamma_{p+q,1}\gamma_{p,0}\right] &=& -E_{p,q}\gamma_{p+q,1}
\gamma_{p,0} + \frac{U}{2M}\,m(p,q)\sum_{k} \,m(k,q)\left(
\gamma_{k+q,0}^{+} \gamma_{k,1}^{+} + \gamma_{k+q,1}
\gamma_{k,0}\right) \nonumber \\
     & & -\frac{U}{2M}\,l(p,q)\sum_{k}\,l(k,q) \left(
\gamma_{k+q,0}^{+} \gamma_{k,1}^{+} -
\gamma_{k+q,1}\gamma_{k,0}
      \right) \nonumber \\
     & &  + \frac{U}{2M}\,n(p,q)\sum_{k}\,n(k,q) \left(
\gamma_{k+q,0} ^{+} \gamma_{k,1}^{+} + \gamma_{k+q,1}
       \gamma_{k,0}\right)    \label{eq: eqmotion2} \\
\left[H , \gamma_{k+q,\sigma}^{+}\gamma_{k,\sigma}\right] &=& (E_{k+q} - E_{k})
                \gamma_{k+q,\sigma}^{+}\gamma_{k,\sigma}.
                               \label{eq: eqmotion3}
\end{eqnarray}
We dropped all terms of the form $\gamma^{+}\gamma$
on the right hand side of the first two equations for
reasons to be discussed shortly.
The coherence factors are defined as
\begin{eqnarray}
\label{coherence:factors}
l(k,q) &=& u_{k}u_{k+q} + v_{k}v_{k+q} \nonumber \\
m(k,q) &=& u_{k}v_{k+q} + v_{k}u_{k+q} \nonumber \\
n(k,q) &=& u_{k}u_{k+q} - v_{k}v_{k+q} \nonumber \\
p(k,q) &=& u_{k}v_{k+q} - v_{k}u_{k+q}
\end{eqnarray}

The next step is to diagonalize the above equations by
finding eigenoperators which satisfy
\begin{equation}
[H , \mu^{+}(q)] = \omega(q)\,\mu^{+}(q)  \label{eq: Hmuplus}
\end{equation}
\begin{equation}
[H , \mu(q)] = -\omega(q)\,\mu(q).  \label{eq: Hmu}
\end{equation}
The ``renormalized'' ground state $|\Psi_{0}>$,  which differs from
the mean field BCS ground state $|\Phi_{0}>$ through the inclusion of
zero-point collective excitations, is defined by
\begin{equation}\label{mupsi}
\mu(q)|\Psi_{0}> = 0.
\end{equation}
In addition $\mu^{+}(q)$ acting on $|\Psi_{0}>$
creates an excited state with an
excitation energy $\omega(q)\ge 0$.

We see that (\ref{eq: eqmotion3}) is already diagonal and the action
of the eigenoperator on an initial state
describes the scattering of a Bogoliubov quasiparticle already
present in that state. There are no excitations present in
the ground state, and
$\gamma_{k+q,\sigma}^{+}\gamma_{k,\sigma}|\Psi_{0}> = 0$.
Since we will look at matrix elements of
the equations of motion between $|\Psi_{0}>$
and an excited state, this operator can be ignored in
the subsequent discussion. As stated earlier,
all the terms containing
$\gamma_{k+q,\sigma}^{+}\gamma_{k,\sigma}$
have already been dropped in
(\ref{eq: eqmotion1}) and (\ref{eq: eqmotion2}).

The non-trivial eigenoperators $\mu^{+}(q)$ and $\mu(-q)$ must
therefore be chosen as a linear combination of
$\gamma_{p+q,0}^{+}\gamma_{p,1}^{+}$ and $\gamma_{p+q,1}\gamma_{p,0}$.
Equivalently, we may write
\begin{equation}
\gamma_{k+q,0}^{+}\gamma_{k,1}^{+} = \left[f(k,q)\mu^{+}(q) +
                                     \tilde{f}(k,q)\mu(-q)\right]
                                     \label{eq: 2gammadager}
\end{equation}
\begin{equation}
\gamma_{k+q,1}\gamma_{k,0} = \left[g(k,q)\mu^{+}(q) +
                             \tilde{g}(k,q)\mu(-q)\right]
                                     \label{eq: 3gammadager}
\end{equation}
where the $f$'s and $g$'s have to be determined.
We substitute (\ref{eq: 2gammadager}) and (\ref{eq: 3gammadager})
in the equations of motion (\ref{eq: eqmotion1}) and
(\ref{eq: eqmotion2}), and take matrix elements
between the ground state $|\Psi_{0}>$ and a
state $|\Psi(q)>$ with a single
quantum of excitation with energy $\omega(q)$.
After some simple algebra we obtain
\begin{equation}
\label{matrix:el:1}
\left[\omega(q)-E_{p,q}\right]\,f(p,q) = \,m(p,q) \,Z(q)
+ \frac{1}{2}\,n(p,q)\,\Lambda(q) + \frac{1}{2}\,l(p,q)\,\Gamma( q)
\end{equation}
\begin{equation}
\label{matrix:el:2}
\left[\omega(q)+E_{p,q}\right]\,g(p,q) = -\,m(p,q) \,Z(q)
 -\frac{1}{2}\,n(p,q)\,\Lambda(q) + \frac{1}{2}\,l(p,q)\, \Gamma(q).
\end{equation}
Note that $\tilde{f}$ and $\tilde{g}$ drop out using (\ref{mupsi}), and
$Z(q)$, $\Lambda(q)$ and $\Gamma(q)$ are defined as
\begin{eqnarray}
\label{coll:coord}
Z(q) &=& -\frac{U}{2M}\sum_{k}\,m(k,q)\,\left[f(k,q) + g(k,q)\right] \\
\Lambda(q) &=& -\frac{U}{M}\sum_{k}\,n(k,q)\,\left[f(k,q) + g(k,q)\right] \\
\Gamma(q) &=& -\frac{U}{M}\sum_{k}\,l(k,q)\,\left[f(k,q) - g(k,q)\right]
\end{eqnarray}

Using (\ref{matrix:el:1}) and (\ref{matrix:el:2}) we find
\begin{equation}
f+g = \frac{1}{\omega^{2}(q)-E_{p,q}^{2}} \left[2E_{p,q}\,m(p,q)\, Z(q)
   + E_{p,q}\,n(p,q)\,\Lambda(q) + \omega(q)\,l(p,q)\,\Gamma(q)\right]
\end{equation}
\begin{equation}
f-g = \frac{1}{\omega^{2}(q)-E_{p,q}^{2}} \left[2\omega(q)\,m(p,q)\,Z (q)
 + \omega(q)\,n(p,q)\,\Lambda(q) + E_{p,q}\,l(p,q)\,\Gamma(q)\right]
\end{equation}
Substituting these results in the
expressions for $Z(q)$, $\Lambda(q)$ and $\Gamma(q)$,
we obtain the three coupled equations
\begin{eqnarray}
\left[1 + U\,I_{E,n,n}(q)\right]\Lambda(q) + U\,I_{\omega,n,l}(q)
\,\Gamma(q) + 2U\,I_{E,n,m}(q)\,Z(q) &=& 0 ~\label{eq: 1eqmotion}\\
U\,I_{\omega,n,l}(q)\Lambda(q) + \left[1 + U\,I_{E,l,l}(q)\right]\,
\Gamma(q) + 2U\,I_{\omega,l,m}(q)\,Z(q) &=& 0 ~\label{eq: 2eqmotion}\\
\frac{U}{2}\,I_{E,m,n}(q)\Lambda(q) + \frac{U}{2}\,
I_{\omega,m,l}(q)\, \Gamma(q) +
\left[1 + U\,I_{E,m,m}(q)\right]\,Z(q) &=& 0 ~\label{eq: 3eqmotion}
\end{eqnarray}
We have used (following ref.[\cite{bs}]) the notation
\begin{equation}\label{iabc}
I_{a,b,c}(q) = {1 \over M}\sum_{k}
\frac{a(k,q)b(k,q)c(k,q)}{\omega^{2}(q)-E_{k,q}^{2}}
\end{equation}
where $a$, $b$, and $c$ denote any one of the following
quantities: the coherence factors defined in (\ref{coherence:factors}), the
quasiparticle energy $E_{k,q}$ defined in (\ref{qp:energy}),
or the excitation energy $\omega(q)$.

The collective mode spectrum $\omega(q)$ is obtained
by solving the secular equation
\begin{equation}\label{det}
\left|
\begin{array}{ccc}
\left[1 + U\,I_{E,n,n}(q)\right] & U\,I_{\omega,n,l}(q)
 & 2U\,I_{E,n,m}(q)
\\
U\,I_{E,n,m}(q) & \left[1 + U\,I_{E,l,l}(q)\right]
 & 2U\,I_{\omega,l,m}(q)
\\
(U/2)\,I_{E,m,n}(q) & (U/2)\,I_{\omega,m,l}(q)
 & \left[1 + U\,I_{E,m,m}(q)\right]
\end{array}
\right|
=
0
\end{equation}
Before solving this equation analytically in several limiting cases,
we can check a general feature of symmetry breaking.
On general grounds we expect a Goldstone mode, i.e., a
solution $\omega(q) = 0$ corresponding to $q = 0$.
(We shall show later that by including the long ranged Coulomb
interaction, this mode will become massive via the Anderson-Higgs mechanism).
By noticing from (\ref{iabc}) that as $q\rightarrow 0$ and $\omega \rightarrow
0$,
$I_{\omega,m,l}$ and
$I_{\omega,n,l}$ vanish, while $I_{E,n,n}$,$I_{E,m,m}$, and $I_{E,n,m}$
remain finite, the secular equation reduces to
\begin{equation}
1 + U\,I_{E,l,l}(0) = 1 - \frac{U}{2M}\sum_{k}\frac{1}{E_{k}} = 0
\end{equation}
This is identical to our gap equation (\ref{gapeqn}), thus providing
a nontrivial check of the consistency of our results.

\medskip
\section{ Weak Coupling}

In weak coupling $\Delta$ is exponentially small and $\tilde{\mu}$ tends to
the fermi energy $\varepsilon_{F}$. The integrals
(\ref{iabc}) are then peaked at $\varepsilon_{F}$.
We use the particle-hole symmetry of the BCS limit to
simplify the calculation.
For $q = 0$, the products of
coherence factors $n(k,q)\,l(k,q)$ and $n(k,q)\,m(k,q)$
are odd under change of sign of $\xi_{k}$, which leads to vanishing
integrals for $I_{\omega,n,l}$ and $I_{E,n,m}$.
As a result we are left with the $2 \times 2$ determinant:
\begin{equation}\label{2by2}
\left| \begin{array}{cc}
1 + U\,I_{E,l,l}(q) & 2U\,I_{\omega,l,m}(q) \\
(U/2)\,I_{\omega,l,m}(q) & 1 + U\,I_{E,m,m}(q)
                                          \end{array} \right| = 0
\end{equation}

The small $q$ and $\omega$ expansion of the various terms
in (\ref{2by2}) is conveniently written in terms of four quantities:
$x=M^{-1}\sum_{k}E_k^{-3}$,
$y=M^{-1}\sum_{k}(\nabla_{k}\xi)^{2}/E_{k}^{3}$,
$w=M^{-1}\sum_{k}\xi\,\nabla_{k}^{2}\xi/E_{k}^{3}$, and
$z=\Delta^{2}M^{-1}\sum_{k}(\nabla_{k}\xi)^{2}/E_{k}^{5}$,
and is given to leading order by
\begin{equation}\label{ell}
1 + U\,I_{E,l,l}(q) = \frac{U}{8}\left[\frac{1}{d}(3z+w-y)q^{2} -
                      x\omega^{2}\right]
\end{equation}
\begin{equation}\label{emm}
1 + U\,I_{E,m,m}(q) = 1 - \frac{U}{2}\Delta^{2}\,x
    \label{eq:emsquare}
\end{equation}
\begin{equation}\label{olm}
U\,I_{\omega,l,m}(q) = (\frac{U}{4}\Delta\,x)\omega
    \label{eq:nulm}
\end{equation}
Substituing these results into
(\ref{2by2}), we obtain the long wavelength dispersion relation
\begin{equation}\label{weakom}
\omega = \left[\frac{1}{d}\,
               \frac{(1 - (U/2)\Delta^{2}x)(3z+w-y)}{x}\right]^{1/2}\,q
\end{equation}
where $d$ is the dimensionality.

The momentum sums are peaked about the Fermi level and may be
estimated by integrals over a thin shell of thickness $2\omega_{c}$,
such that $\Delta\ll \omega_{c}\ll W=2dt$,
centered around the Fermi energy.
We then find
$x = 2N(0)/\Delta^{2}$,
$y = 2N_{\rm v}(0)/\Delta^{2}$,
$w = -2a^{2}N(0)\log(2\omega_c/\Delta)$, and
$z = 4N_{\rm v}(0)/3\Delta^{2}$,
where the density of states
$N(\xi) =  {(2\pi)^{-d}} \int d^{d}k \delta(\xi_k - \xi)$
and
\begin{equation}\label{nsubv}
N_{\rm v}(\xi) = \int \frac{d^{d}k}{(2\pi)^{d}}
(\nabla_{k}\xi_{k})^{2} \delta(\xi_k - \xi).
\end{equation}
The only dependence on the (arbitrary) cutoff $\omega_c$ is in $w$,
which, however, is negligible compared to the other terms in
(\ref{weakom})
in the weak coupling limit where $\Delta$ is small. (In the continuum
limit $w$ is identically zero).

The weak coupling, long wavelength dispersion is thus given by
\begin{equation}\label{weakom1}
\omega(q) = \left[\langle v^2_F \rangle /d \right]^{1/2} (1 - UN(0))^{1/2}\ q
\end{equation}
where the mean squared Fermi velocity is defined by
$\langle v^2_F \rangle \equiv N_{\rm v}(0) / N(0)$.

The density, or filling, dependence of $\langle v^2_F \rangle$ depends upon the
band structure. For the tight binding model used here, we obtain
\begin{equation}
\frac{\langle v^2_F \rangle}{(2ta)^2}
                  = \frac{\int d^{d}k (\sum_{i=1}^{d}\sin^{2}(k_{i}a))
  \delta(\xi_{k})}{\int d^{d}k \delta(\xi_{k})}
\end{equation}
and the dependence of $\langle v^2_F \rangle$ on the filling $f$ is plotted in
Fig.~5.
In general the $f$-dependence
is smooth  with the following exceptions. In the 3D case there is
a sharp dip in the collective mode speed of sound at $f = 0.125$
due to a van Hove singularity. In the 2D case the speed of sound goes to
zero as $f \to 1/2$ due to the nesting at half-filling.

In the continuum limit, with a parabolic dispersion,
$\sqrt{\langle v_F^2 \rangle} = p_F/m$
and (\ref{weakom1}) reduces to Anderson's weak coupling result\cite{pwa} in
3D.  The collective mode in weak coupling is essentially the same for the
lattice and continuum models, as one might expect since the
size of the bound pairs $\xi_0$ is much larger than the lattice spacing $a$.

\vspace{0.5cm}
\section{Strong Coupling}

In the strong coupling limit the fermions bind into on-site
singlet pairs, and the low energy physics of the
attractive Hubbard model can be obtained by a projection
on to the non-singly occupied subspace.
This exact mapping\cite{rmp} results in
a system of hard core bosons described by
\begin{equation}
H_{\rm bose} = \sum_{i,j} \left(\frac{2\,t^{2}}{U}\right)\,(n_{i}\,n_{j}
                                                  - b_{i}^{+}b_{j}).
\end{equation}
The composite bosons move only via virtual ionization with
an effective hopping amplitude $t_{\rm b}=-2t^{2}/U$.
The hard core constraint is due to the Pauli
principle for the constituent fermions, and in addition
the bosons interact with a nearest neighbor repulsion $V=2t^{2}/U$.
In the long wavelength limit, the collective
excitation of a dilute ($na_s^3 \ll 1$)
3D Bose gas is the Bogoliubov sound mode\cite{bog} with dispersion
$\omega = \left(4\pi n_{\rm b} a_s/m_{\rm b}^2\right)^{1/2}q$.
The density of bosons is $n_{\rm b} = f/a^{3}$
and the effective mass is $m_{\rm b} \simeq -1/t_{\rm b}a^2 = U/2t^2a^2$.
In terms of the parameters $f$, $a$, $U$ and $t$ which characterize the
constituent fermions, the Bogoliubov dispersion is given by
\begin{equation}\label{bogsgom}
\omega(q) \simeq \sqrt{\pi f}\,(4t^2a/U)\,q ,
\end{equation}
where we take the scattering length $a_s$,
describing the interactions between
the bosons, to be of the order of the lattice spacing $a$.

We now show that one obtains essentially the same result from
strong coupling limit of the general expression (\ref{det}), which comes
from an RPA analysis of the constituent fermions.
Using the gap and the number equations, (\ref{gapeqn}) and
(\ref{numbereqn}),
we find
$\Delta = U[f(1-f)]^{1/2}\left[1 - d\alpha^2\right]$
and
$\tilde{\mu} = U(2f-1)\left[1 + 2d\alpha^2\right]/2$,
to leading order in $\alpha = 2t/U \ll 1$.
A similar expansion of the various quantities in (\ref{det})  (see appendix
A for details of derivations), yields
\begin{eqnarray}
1 + U\,I_{E,n,n} &=& 4f(1-f)\left[1 + 4(1-6f+6f^{2})d\alpha^2 \right]
                \label{1uienn}\\
1 + U\,I_{E,m,m} &=& (2f-1)^{2} + 8f(1-f)\left[1 - 3(2f-1)^{2}\right]d
                                                 \alpha^2 \\
U\,I_{E,n,m} &=& (2f-1)[4f(1-f)]^{1/2}\left[1 - (5-6(2f-2)^{2})d
                                                 \alpha^2 \right] \\
U\,I_{\omega,n,l} &=& \frac{\omega\,(2f-1)}{U}\left[1 - 2(2-3(2f-1)^{2})d
                                                  \alpha^2\right] \\
U\,I_{\omega,m,l} &=& \frac{\omega}{U}[4f(1-f)]^{1/2}\left[-1 +
(1+6(2f-1)^{2})d
                                                  \alpha^2 \right]     .
\end{eqnarray}
\noindent
We also find, independent of the filling,
\begin{equation}
1 + U\,I_{E,l,l} = \alpha^2q^2a^2/2\,- \omega^2/U^2.
\end{equation}
With the further simplification of the dilute limit $f \ll 1$, the
various quantities become:
$1 + U\,I_{E,n,n} = 4f\left[1 + 4d\alpha^2\right]$,
$1 + U\,I_{E,m,m} = 1 - 16fd\alpha^2$,
$U\,I_{E,n,m} = -2f^{1/2}\left[1 +d\alpha^2\right]$,
$U\,I_{\omega,n,l} = -\omega\left[1 + 2d\alpha^2\right]/U$, and
$U\,I_{\omega,m,l} = 2\omega f^{1/2}\left[-1 + 7d\alpha^2 \right]/U$.
We then obtain the RPA result for the dilute ($f \ll 1$),
strong coupling ($U/t \gg 1$) limit
\begin{equation}\label{rpasgom}
\omega(q) = \sqrt{4df}\,(4t^2a/U)\,q
\end{equation}
in $d$-dimensions.
This result is essentially the same as that expected for the dilute
Bose gas (\ref{bogsgom}). The RPA gets both the square root density dependence
and
the inverse boson mass dependence of the speed of sound correctly. The
only difference is an overall factor of the order of unity, which is
not unexpected since we had simply used $a_s \sim a$ in (\ref{bogsgom}).
Quite remarkably,
starting with interacting fermions and using RPA we were able to reach
the regime of hard core bosons in the strong coupling limit.

\section{Crossover}

Encouraged by the success of the RPA in strong coupling,
we might expect it to be a reasonable interpolation scheme
all the way from weak to strong coupling.
For intermediate couplings we have numerically solved the RPA
equation (\ref{det}) as a function of the coupling
$U/t$, and for various fillings, $f=0.05, 0.25, 0.45$.
As input to these equations we have used the numerical solutions
for $\Delta$ and $\tilde{\mu}$ obtained from the mean field gap and number
equations, (\ref{gapeqn}) and (\ref{numbereqn}); (for the 2D case see Fig.~1).
To evaluate the $k$-sums in (\ref{det}) we used slightly different
procedures in 2D and 3D. In three dimensions we numerically evaluated
the density of states $N(\xi)$ and the function
$N_{\rm v}(\xi)$ (see (\ref{nsubv})) for the simple cubic
nearest neighbor tight binding band structure.
In 2D we used analytical expressions, in terms of complete elliptic integrals,
for these density of states functions for the square lattice.
(See appendix B for derivation of $N_{\rm v}(\xi)$ in 2D).

The collective mode velocity
is plotted in Fig.~4a (for 2D) and Fig.~4b (for 3D)
as a function of the coupling.
These results supercede those given in Fig.~2 of our previous
paper\cite{br} which contained a numerical error.
The analytical results in the  small
$U$, obtained from (\ref{weakom1}), are separated for clarity from the
numerical curves in Fig.~4a and 4b. In strong coupling, the $f=0.05$
result (\ref{rpasgom}), is compared with the numerical result at the same
filling.
The $f=0.45$ numerical result is compared in strong coupling with the
analytical result for half filling $f=0.5$ which is much easier to obtain.
We see that at all fillings, the numerical results smoothly interpolate
between the analytical weak and strong coupling results.
The $1/U$ dependence of $c$ is apparent in
strong coupling, at all fillings. Note also the
non-monotonic dependence on $f$ in weak coupling
as discussed above.

\section{Charged Systems}

We now extend our analysis
to account for the effects of the long ranged Coulomb interaction
between fermions.
It is well known that in the BCS limit
the singular behavior of the Coulomb interaction in
the long wavelength limit has a dramatic effect on the
collective modes. Anderson first showed\cite{pwa} that in three dimensions,
within the weak coupling  approximation the sound mode is pushed
up to the plasma frequency which is well above the energy gap.
This showed that the collective modes were in fact quite unimportant for
charged superconductors, except to restore gauge
invariance in BCS theory\cite{parks}. However as the coupling increases,
we expect the energy gap to increase and the plasma frequency to
decrease due to the increasing effective mass of the bound pairs
in the strong coupling lattice model.
In the limit of very large $U$ the spectrum
should tend to the plasma frequency of a charged bose gas\cite{foldy}
$\omega={\sqrt{4\pi\,n_{\rm B}e^{2}/m_{\rm B}}}$ which goes
to zero as $U$ goes to infinity, since the mass of the boson
$m_{\rm B} \sim U/t^2$.

The inclusion of the Coulomb interaction in the negative-$U$ Hubbard model
is necessarily somewhat ad-hoc. The Negative-$U$ Hubbard Model
is only an effective model
where $-U$ can be viewed as the screened interaction,
and where the bare Coulomb interaction  $V_{c}(q)$
is absent. In the RPA, screening is accomplished by attaching to each
vertex a series of the polarization vertex shown in Fig.~3a.
We see that the RPA equations of motion automatically screen
the interaction vertex in the processes shown in Fig.~2c and 3a.
Therefore we can
use the bare interaction $V_{c}(q)$ in these two terms.
However the interaction vertices in the remaining processes
in Fig.~2 and 3 cannot be connected to the polarization
vertex and thus are not screened within RPA.
We must therefore use the screened interaction, namely
$-U$, for these terms.
This procedure has the effect of coupling the bare interaction
$V_{c}$ to the particle-hole channel only, where
$V_{c}$ is given by
\begin{equation}
V_{c}(q)= {2(d-1)\pi\,e^{2} \over a^d q^{d-1}}
\end{equation}
and $a^d$ is the unit cell volume in $d$-dimensions.

Incorporating these modifications in the equations of motion for
$c^\dagger c^\dagger$ and $c^\dagger c$, and transforming to the
Bogoliubov representation, we find that
the equations of motion (\ref{det}) are modified in the following way
\begin{equation}\label{detch}
\left[1 + U\,I_{E,n,n}(q)\right]\Lambda(q) + U\,I_{\omega,n,l}(q)
\,\Gamma(q) + 2U\,I_{E,n,m}(q)\,Z(q) = 0
\end{equation}
\begin{equation}
U\,I_{E,n,m}(q)\Lambda(q) + \left[1 + U\,I_{E,l,l}(q)\right]\,
\Gamma(q) + 2U\,I_{\omega,l,m}(q)\,Z(q) = 0
\end{equation}
\begin{equation}
-V_{c}(q)\,I_{E,m,n}(q)\Lambda(q) - V_{c}(q)\,I_{\omega,m,l}(q)\,
\Gamma(q) + \left[1 - 2V_{c}(q)\,I_{E,m,m}(q)\right]\,Z(q) = 0
\end{equation}
The reason why $V_c$ enters only in the last row of (\ref{det}),
replacing the factors of $U/2$, can be traced to
the collective coordinate $Z$ of (\ref{coll:coord})
which involves the coherence factor $m = uv' + vu'$
and is thus related to the p-h channel.

We will restrict ourselves here to the dense limit, i.e., close
to half filling. The dilute regime is more complicated, since
we know that, in the strong coupling limit, we would obtain a system
of charged bosons, which at sufficiently low densities, would form a
Wigner crystal. Thus one would expect a first order phase transition
to a crystal with decreasing density, which lies beyond the scope of
the present work.
To simplify the algebra we will work at half-filling $f=0.5$
assuming a state with off-diagonal long ranged order (and
ignoring CDW ordering). Since ${\tilde{\mu}} = 0$ for $f = 0.5$
independent of $U$, there is an exact particle-hole symmetry
for all $U$ which greatly simplifies the calculation.
Thus $I_{\omega,n,l} = I_{E,n,m} = 0$ and
we are left with the 2$\times$2 determinant.
\begin{equation}
\left| \begin{array}{cc}
1 + U\,I_{E,l,l}(q) & 2U\,I_{\omega,l,m}(q) \\
-V_{c}(q)\,I_{\omega,l,m}(q) & 1 - 2V_{c}(q)\,I_{E,m,m}(q)
                                          \end{array} \right| = 0
\end{equation}
The collective modes are obtained by solving the equation
\begin{equation}\label{disp:charged}
1 = 2V_{c}(q)\left[ I_{E,m,m} - U\frac{I_{\omega,l,m}^{2}}{1+UI_{E,l,l}}\right]
\end{equation}

It is convenient to write
$I_{a,b,c}(q) = I_{a,b,c}^{0} + \delta\,I_{a,b,c}(q)$,
where $I_{a,b,c}^{0} = I_{a,b,c}(q=0)$.
Using
\begin{equation}
I_{E,m,m}^{0} =
{1\over M}\sum_{k}\frac{2\,\Delta^{2}}{E_{k}(\omega^{2}-4E_{k}^{2})}
\end{equation}
\begin{equation}
I_{\omega,l,m}^{0} =
{1\over M}\sum_{k}\frac{\omega\Delta}{E_{k}(\omega^{2}-4E_{k}^{2})}
\end{equation}
\begin{equation}
I_{E,l,l}^{0} =
{1\over M}\sum_{k}\frac{2\,E_{k}}{\omega^{2}-4E_{k}^{2}}
\end{equation}
and noticing that
$I_{\omega,l,m}^{0} = \left(2\Delta/U\omega\right)(1 + U\,I_{E,l,l}^{0})$,
it is straightforward to show, using the gap equation (\ref{gapeqn}), that
\begin{equation}
I_{E,m,m}^{0} - U\frac{(I_{\omega,l,m}^{0})^{2}}{1 + U\,I_{E,l,l}^{0}}
            = 0.
\end{equation}
The dispersion relation then simplifies to
\begin{equation}\label{chdisp3D}
1 = 2V_{c}(q)\left[\delta\,I_{E,m,m}(q) - \frac{4\Delta}{\omega}
\delta\,I_{\omega,l,m}(q) + \frac{4\Delta^{2}}{\omega^{2}}
\delta\,I_{E,l,l}(q)\right]
\end{equation}
We will study now the limiting case of weak and strong coupling
in the 3D and 2D cases.

\indent
\subsection{3D Weak coupling}

We expect here the collective mode to be pushed up to a plasma
frequency much higher than the energy gap. The dominant contribution
is given by $\delta\,I_{E,m,m}(q)$, which after some algebra can be
written in the form
\begin{equation}
\delta\,I_{E,m,m}(q) = \frac{q^{2}}{3\omega^{2}}\frac{3\Delta^{2}}{2}
         {1\over M} \sum_{k}\left(\frac{\xi_k^{2}(\nabla_\xi)^{2}}{E_k^{5}} -
          \frac{\xi_k(\nabla_k^{2}\xi_k)}{E_k^{3}}\right)
\end{equation}
Using the same tricks that were used in the weak coupling analysis
of the neutral case (below equation (\ref{weakom})), we see that
the second term on the right hand side can
be neglected for $\Delta \rightarrow 0$ and the first
term gives the result
\begin{equation}
\delta\,I_{E,m,m}(q) = \frac{q^{2}}{3\omega^{2}}N_{\rm v}(0).
\end{equation}
Substituting this into (\ref{chdisp3D}) we find
\begin{equation}\label{3dcweakom}
\omega^{2} = \frac{8\pi}{3}e^{2}\langle v^2_F \rangle N(0),
\end{equation}
which reduces to the plasma frequency
$\omega^2 =  4\pi\,ne^{2}/m$
for a parabolic dispersion.

\indent
\subsection{3D Strong coupling}

In the strong coupling regime $\Delta \gg \omega$, and the term
$(4\Delta^{2}/\omega^2)\delta\,I_{E,l,l}$ dominates
eq.\ (\ref{chdisp3D}).
After a small $q$ expansion of $\delta\,I_{E,l,l}$
the dispersion relation can be written as
\begin{equation}
1 =V_{c}(q)\frac{\Delta^{2}q^{2}}{3\omega^{2}}{1\over M}\sum_{k}
   \frac{(\nabla_k\xi_k)^{2}}{E_k^{3}}.
\end{equation}
Using the explicit form of $V_{c}(q)$ we then obtain
\begin{equation}\label{3dcstrongom}
\omega^{2} = \frac{4\pi\,e^{2}}{a^{3}}( \frac{4t^{2}a^{2}}{U})
\end{equation}
Using the boson mass $m_{\rm B} = U/2t^2a^2$ (as in the neutral case),
charge $e_{\rm B} = 2e$, and density $n_{\rm B} = 1/2a^{3}$
(since the constituent fermions are close to half filling),
we can rewrite the above result as
\begin{equation}
\omega^{2} = \frac{4\pi\,n_{\rm B}\,e_{\rm B}^{2}}{m_{\rm B}}
\end{equation}
This is exactly the plasma frequency
of a dense system of charged bosons\cite{foldy}, thus
providing another important check of
the validity of RPA in the strong coupling regime at $T=0$.

\indent
\subsection{Collective modes in 2D}

In two dimensions, $V_{c}(q)= {2\pi\,e^{2}}/{a^2q}$, and we expect the
plasma frequency $\omega_p \sim \sqrt{q}$ as $q\rightarrow 0$.
To calculate the collective mode spectrum, one can then use a small $\omega$
and
small $q$ expansion of (\ref{disp:charged}). The calculation proceeds
in a manner analogous to the neutral system calculation, and
we find the dispersion relation
\begin{equation}\label{chdisp2D}
\frac{1}{d}V_c\,\Delta^2 x(3z + w - y)q^2 = x\omega^2
\end{equation}
where we use the quantities $x,y,z,w$ introduced below (\ref{2by2}).

In the BCS limit we use
the expressions for $x$,$y$,$w$ and $z$ derived earlier
for the weak coupling neutral case (see below {\ref{weakom}))
and obtain the result
\begin{equation}\label{2dweakom}
\omega^{2} = 2\pi\,e^{2}\langle v^2_F \rangle N(0)\,q
\end{equation}
For comparison, it might be useful to look at the continuum results
where $N(0)=m/2\pi$, $\langle v^2_F \rangle=(p_F/m)^2$, and $n= 2\pi p_F^2$
so the (\ref{2dweakom}) can be written as
$\omega = \left(2\pi e^2n/m\right)^{1/2}\,\sqrt{q}$.

In the strong coupling Bose limit
one can use either (\ref{chdisp2D}), or go back to (\ref{chdisp3D})
with the additional simplification that $\omega \ll \Delta$.
Both routes yield the same result
\begin{equation}\label{2dstrongom}
\omega^{2} = \left({{8\pi e^2 t^2}\over{U}} \right)\,q
\end{equation}
This may be written in terms of the bose parameters
$m_{\rm B} = U/2t^2a^2$, $n_{\rm B} = 1/2a^2$, and $e_{\rm B} = 2e$
to obtain
$\omega= \left(2\pi e_{\rm B}^2 n_{\rm B}/m_{\rm B}\right)^{1/2}\,\sqrt{q}$.

We have numerically calculated the crossover in the 2D plasmon mode
from the BCS regime to the Bose regime
close to half-filling. The results are plotted in
Fig.~6: we find a monotonically decreasing function of $U$
which smoothly interpolates between the
analytical results in weak (\ref{2dweakom}) and strong coupling
(\ref{2dstrongom}).

\medskip
\section{Conclusion}

In conclusion
we have studied the attractive Hubbard model in two and three
dimensions which is the simplest lattice model
that shows a crossover from weak coupling BCS theory to
a strong coupling Bose condensed regime.
Using a generalized RPA formulation for the collective excitations
at $T=0$ we were able to analytically reproduce the known results
in the BCS and the Bose limits, for both neutral and charged systems,
and numerically show that
there was a smooth evolution from one limit to the other.
This suggests that the RPA provides a reasonable
interpolation scheme in the intermediate regime.

Several open questions remain, some of which are listed below.
Generalization to layered superconductors should be of interest.
Collective modes in layered systems
in the weak coupling BCS regime have been studied recently\cite{fds}
and a formalism to study the crossover problem in layered superconductors
has been developed by Cote and Griffin\cite{griffin}.
The generalization to finite temperatures would also be of
considerable interest.
Recently Sa de Melo, Engelbrecht and one of the authors have
used the functional integral method to address this problem in a continuum
model\cite{sre}. The lattice model has some important differences from
the continuum problem: in particular $T_c$ for the attractive Hubbard
model is expected to grow in the BCS
manner like $\exp(-t/U)$  for weak coupling but eventually drop like\cite{rmp}
$t^2/U$ in the Bose regime. It would be very interesting to
derive a general expression for $T_c(U/t)$ and to determine its
maximum value which necessarily occurs in the intermediate coupling regime.

\acknowledgements

M. R. would like to acknowledge support by the U.S. Department
of Energy, Office of Basic Energy Sciences under contract
no.~W-31-109-ENG-38.

\newpage
\setcounter{section}{0}
\section{Appendix A}

\subsection{Part I}

In the first part of this appendix we shall calculate
the quantities $\tilde{\mu}$,
$\Delta$, and $E_0 = (\tilde{\mu}^{2}+\Delta^{2})^{1/2}$
in strong coupling
to order $\left(t/U\right)^2$ at arbitrary filling.
We will often use the identities
\begin{equation}\label{identity1}
\sum_{k}\varepsilon_{k} = 0
\end{equation}
and
\begin{equation}\label{identity2}
{1\over M}\sum_{k}\varepsilon_{k}^{2} = 2dt^2
\end{equation}

We start by expanding the gap equation (\ref{gapeqn})
to second order in $\varepsilon_{k}/E_0$, which is
at most of order $t/U$ in the strong coupling limit:
\begin{eqnarray}
1 &=& \frac{U}{2M}\sum_{k}\frac{1}{E_0}
\left[1 - \frac{\varepsilon_{k}^{2}-2\varepsilon_{k}\tilde{\mu}}
{2E_0^{2}} + \frac{3\varepsilon_{k}^{2}\tilde{\mu}^{2}}{2E_0^{4}}
\right]  \nonumber \\
  &=& \frac{U}{2E_0}\left[1 - d(\frac{1}{4E_0^{2}} +
      \frac{3d\tilde{\mu}^{2}}{4E_0^{2}})(\frac{2t}{U})^{2}\right]
\nonumber
\end{eqnarray}
This can be solved for $E_0$ to yield
\begin{equation}
E_0 = \frac{U}{2}\left[1 +
d\left(3(\frac{2\tilde{\mu}}{U})^{2}-1\right)(\frac{2t}{U})^{2}\right]
\label{eq: Ezero}
\end{equation}
Now expanding the number equation
\begin{equation}
2f = 1 + \frac{2\tilde{\mu}}{U} -
\sum_{k}\frac{\varepsilon_{k}}{E_0}\left[1 -
\frac{\varepsilon_{k}^{2}-2\varepsilon_{k}\tilde{\mu}}{2E_0^{2}} +
\frac{3\varepsilon_{k}^{2}\tilde{\mu}^{2}}{2E_0^{4}}\right] \\
\nonumber
\end{equation}
rearranging the equation and using the leading term of $E_0$
expansion in the third term of the right hand side
\begin{equation}
2f - 1 =  \frac{2\tilde{\mu}}{U}\left[1 - 2d(\frac{2t}{U})^{2}\right]
\nonumber
\end{equation}
Again by inverting the equation, we find
\begin{equation}
\tilde{\mu} = \frac{U}{2}(2f-1)\left[1 + 2d(\frac{2t}{U})^{2}\right]
\label{eq: mutilde}
\end{equation}
Substituing (~\ref{eq: mutilde}) in (~\ref{eq: Ezero}) we find
\begin{equation}
E_0 = \frac{U}{2}\left[1 + d\left(3(2f-1)^{2}-1\right)
         (\frac{2t}{U})^{2}\right]
\label{eq: Ezero2}
\end{equation}
using $\Delta^{2}=E_0^{2}-\tilde{\mu}^{2}$, and
eqs.(~\ref{eq: mutilde},~\ref{eq: Ezero2}) we find
\begin{equation}
\Delta =
\frac{U}{2}\left[4f(1-f)\right]^{1/2}\left(1-d(\frac{2t}{U})^{2}\right)
\label{eq: delta}
\end{equation}

\subsection{Part II}

Here we derive in detail the strong coupling
expressions of the quantity $1 + U\,I_{E,n,n}$; the derivation of the
other coefficients in the equations of motion follow is very similar.
Since this quantity has a non-vanishing limit as $q\rightarrow 0$
we can simply set $q=0$ at the outset, thus obtaining
\begin{eqnarray}
1 + U\,I_{E,n,n} &=& 1 -
\frac{U}{2M}\sum_{k}\frac{(\varepsilon_{k}-\tilde{\mu})^{2}}
{\left[(\varepsilon_{k}-\tilde{\mu})^{2} + \Delta^{2}\right]^{3/2}}
\nonumber \\
         &=& 1 - \frac{U}{2M}\sum_{k}\frac{\varepsilon_{k}^{2} +
\tilde{\mu}^{2} - 2\varepsilon_{k}\tilde{\mu}}
{E_0^{3}}\left[1-\frac{3}{2}\frac{\varepsilon_{k}^{2}-
2\varepsilon_{k}\tilde{\mu}}{E_0^{2}}+
\frac{15\varepsilon_{k}^{2}\tilde{\mu}^{2}}{2E_0^{4}}\right]
\nonumber
\end{eqnarray}
Using the property $\sum_{k}\varepsilon_{k} = 0$ we find
\begin{eqnarray}
1 + U\,I_{E,n,n} &=& 1 - \frac{U\tilde{\mu}^{2}}{2ME_0^{3}}
- \frac{U}{2ME_0^{3}}\sum_{k}\varepsilon_{k}^{2}
(1 - \frac{15\tilde{\mu}^{2}}{2E_0^{2}} +
\frac{15\tilde{\mu}^{4}}{2E_0^{4}}) \nonumber \\
                 & &            \nonumber
\end{eqnarray}
Using eqs(~\ref{eq: mutilde},~\ref{eq: Ezero2}) we get
\begin{eqnarray}
1 + U\,I_{E,n,n} &=& 1 - \left[1 - 3\left(3(2f-1)^{2}-1\right)
d(\frac{2t}{U})^{2}\right]\times \nonumber \\
                 & & \left[(2f-1)^{2}\left(1+4d(\frac{2t}{U})^{2}
\right) + \left(2-15(2f-1)^{2}+15(2f-1)^{4}\right)d(\frac{2t}{U})^{2}
\right]  \nonumber \\
                 &=& 4f(1-f)\left[1 + 4(1-6f+6f^{2})d
                   (\frac{2t}{U})^{2}\right] \label{eq: Ensquare}
\end{eqnarray}
Equation (~\ref{eq: Ensquare}) is identical to eq.\ref{1uienn}.

\medskip
\section{Appendix B}

We will derive here the expression of the weighted density of states
in 2-D defined by
\begin{equation}
 N_{\rm v}(\varepsilon) = \int \frac{d^{2}k}{(2\pi)^{2}}
(\nabla\varepsilon_{k})^{2} \delta(\varepsilon - \varepsilon_{k})
\end{equation}
This can be extracted from the advanced green's
function\cite{economou} using the relation
\begin{equation} \label{eq:ro-greenfun}
N_{\rm v}(\varepsilon) = \frac{1}{\pi}Im(G_{1}^{+}(\varepsilon))
\end{equation}
\begin{eqnarray}
G_{1}^{+}(z) &=& \frac{1}{(2\pi)^{2}}\int d^{2}k
\frac{(\nabla\varepsilon_{k})^{2}}{z-\varepsilon_{k}} \nonumber \\
             &=& \frac{(2ta)^{2}}{(2\pi)^{2}}\int_{-\pi}^{\pi}
dk_{1}\int_{-\pi}^{\pi}dk_{2}\frac{\sin^{2}k_{1}+\sin^{2}k_{2}}
{z - 2t(\cos k_{1}+\cos k_{2})}   \label{eq: grenndef} \\
             &=& \frac{(Wa)^{2}}{(2\pi)^{2}}\int_{-\pi}^{\pi}
d\phi_{1}\int_{-\pi}^{\pi}d\phi_{2}\frac{\cos^{2}\phi_{1}\sin^{2}
\phi_{2}}{z - W\,\cos\phi_{1}\cos\phi_{2}} \label{eq: moritatransf}
\end{eqnarray}
for a derivation of (\ref{eq: moritatransf}) see ref\cite{morita}. We
have now the product of two simple integrals which can be put in the
form
\begin{equation}
G_{1}^{+}(z) = \frac{(Wa)^{2}}{(2\pi)^{2}}\int_{-\pi}^{\pi}d\phi_{1}
\cos^{2}\phi_{1}\frac{I(\phi_{1})}{W\,\cos\phi_{1}} \label{eq: GzIfi}
\end{equation}
where $W=4t$ is half the bandwidth, $a$ is the lattice spacing, and
\begin{equation}
I(\phi_{1}) = \int_{-\pi}^{\pi}d\phi\frac{\sin^{2}\phi}{x - \cos\phi};
                         \hspace{2cm}  x = \frac{z}{W\,\cos\phi_{1}}
\label{eq: Ifi}
\end{equation}
A contour integral evaluation of the integral above leads to
\begin{equation}
I(\phi_{1}) = \frac{-2\pi}{x + \sqrt{x^{2}-1}} \label{eq:
Ifiintegrated}
\end{equation}
Substituing (\ref{eq: Ifiintegrated}) into (\ref{eq: GzIfi}) we find
\begin{eqnarray}
G_{1}^{+}(z) &=& -z\left[1 - \frac{2}{\pi}\int_{0}^{\pi/2}
d\phi\sqrt{1-(W/z)^{2}cos^{2}\phi}\right] \nonumber \\
             &=& -z\left[ 1 - \frac{2}{\pi}{\bf E}(W/z) \right]
\label{eq: finalG1}
\end{eqnarray}
where $\bf{E}$ is the complete elliptic integral of the second kind.
In the case where $W/|\varepsilon| \leq 1$ (i.e $\varepsilon$
is taken within the
band), one can use the analytic continuation of $\bf{E}$ given
by\cite{morita}
\begin{equation}
{\bf E}\left(\frac{1}{\varepsilon}\right) = \frac{1}{\varepsilon}
\left[{\bf E}(\varepsilon)
-i{\bf E}(\sqrt{1-\varepsilon^{2}}) -
\varepsilon^{2}{\bf K}(\varepsilon) +
i\varepsilon^{2}{\bf K}(\sqrt{1-\varepsilon^{2}})\right]
\label{eq: 2ellip}
\end{equation}
where ${\bf K}$ is the complete elliptic integral of the first kind,
and $\varepsilon$ is expressed in units of $W$.
Taking the imaginary part (\ref{eq: 2ellip}), substituing it
into (\ref{eq: finalG1}) and using (\ref{eq:ro-greenfun}) we
finally find
\begin{equation}
N_{\rm v}(\varepsilon) =
\frac{2}{\pi^{2}}\left[ {\bf E}(\sqrt{1-\varepsilon^{2}})
- \varepsilon^{2}{\bf K}(\sqrt{1-\varepsilon^{2}})\right]
\label{eq: moddos}
\end{equation}
Equation (\ref{eq: moddos}) was used in the calculation of the energy
gap $\Delta$, the chemical potential $\mu$, as well as the velocity
of the collective mode $c$ in 2D.
For completeness we give also the density of states $N(\varepsilon)$
in 2D\cite{economou}. The figures of $N(\varepsilon)$ and $N_{\rm
v}(\varepsilon)$ for both the 2D and 3D cases are shown in figure 7.
\begin{equation}
N(\varepsilon) = \frac{2}{\pi^2}{\bf K}(\sqrt{1-\varepsilon^{2}})
\end{equation}

\newpage

\newpage
%
%
\begin{figure}
   {\caption{ (a) The order
parameter $\Delta$ and b) the chemical potential $\mu$, as
functions of the coupling $U$ for 2D. All energies are plotted
in units of the half-bandwidth $W=4t$.The dashed, solid and
dotted lines correspond to filling factors $f=
0.45$, 0.25, and 0.05 respectively.
\label{fig1}
 }}
\end{figure}

\begin{figure}
   {\caption{
   Diagrammatic representation of various terms arising to first
order in the interaction in the pair, or particle-particle channel.
The dashed line represents the interaction $U$, and the black circle
represents the condensate.
\label{fig2}
}}

\end{figure}

\begin{figure}
   {\caption{
Diagrammatic representation of various terms arising to first
order in the interaction in the density fluctuation,
or particle-hole channel.
The dashed line represents the interaction $U$, and the black circle
represents the condensate.
\label{fig3}
}}

\end{figure}

\begin{figure}
   {\caption{
Collective Mode velocity as a function of
coupling $U$ for various fillings in (a) two and (b) three dimensions.
$W=2dt$ is half the bandwidth, $d$ the dimensionality, and $a$ the
lattice spacing. The dashed, solid and dotted
lines correspond to $f=$ 0.45, 0.25, and 0.05 respectively.
The results
of the numerical evaluation are separated by breaks in the curve
from the weak and strong coupling results obtained anaytically.
\label{fig4}
}}
\end{figure}

\begin{figure}
   {\caption{
Fermi velocity in 2D and 3D as a function of filling in the weak
coupling limit. The sharp feature near $f=0.21$ for the 3D case is a result
of a van Hove singularity.
\label{fig5}
}}
\end{figure}

\begin{figure}
   {\caption{
Collective Mode velocity  as a function of
coupling $U$ for the 2D charged system in the dense limit
($f\simeq 0.5$). The result of the numerical evaluation is separated by breaks
in the curve from the weak and strong coupling results obtained anaytically."
\label{fig6}
}}
\end{figure}

\begin{figure}
    {\caption{
The density-of-states functions
$N(\varepsilon)$ and $N_{\rm v}(\varepsilon)$ for (a) 2D and (b) 3D
tight binding model.
\label{fig7}
}}
\end{figure}

\end{document}